\documentstyle[twocolumn,aps,epsf]{revtex}
\begin{document}
 
\title{Electronic Spectral Functions for Quantum Hall Edge States}
 
\author{U.\ Z\"ulicke and A.\ H.\ MacDonald}

\address{
Department of Physics, Indiana University, Bloomington, Indiana 
47405, U.S.A.
}

\date{\today}
\maketitle
 
{\tightenlines
\begin{abstract}

We have evaluated wavevector-dependent electronic spectral functions
for integer and fractional quantum Hall edge states using a chiral
Luttinger liquid model. The spectral functions have a finite width
and a complicated line shape because of the long-range of the 
Coulomb interaction. We discuss the possibility of probing these
line shapes in vertical tunneling experiments.

\end{abstract}
}
 
\pacs{PACS numbers: 73.40.Hm, 73.20.Mf, 73.40.Gk, 72.15.Nj} 
 
\narrowtext

The quantum-Hall (QH) effect occurs in high mobility two-dimensional
electron systems (2DES's) under the influence of strong 
perpendicular magnetic fields. Kinetic energy quantization and 
strong correlations between electrons can lead to discontinuities in
the dependence of the chemical potential of the system on density at
zero temperature,  {\it i.e.} to incompressible ground states. 
Although no gap-less charged excitations of incompressible states
can occur in the bulk of the system, the fact that these states 
occur\cite{ahmintro} at magnetic field dependent densities implies
that gap-less charged excitations {\it do} occur at the edge of the
2DES. The physics of edge excitations has played an important role
in explaining transport properties of QH 
systems\cite{rbl:prb:81,bih:prb:82,pav:prb:84,butt:prb:88}. The edge
of a 2DES in the QH effect regime is an interesting and in many
senses an ideal realization of a one-dimensional (1D) Fermion 
system\cite{emery,sol:adv:79,fdmh:jpc:81}. A unique feature is the
spatial separation of the left- and right-moving branches of the
spectrum which is permitted by time-reversal symmetry breaking in a
magnetic field and can make impurity-backscattering negligible,
creating an opportunity to study interaction effects on transport
properties in 1D\cite{apel:prb:82} experimentally.

For a non-interacting 1D Fermion system, the electronic spectral
function has a single unit weight $\delta$-function peak at the
quasiparticle energy, $\xi_k = \pm \hbar v_F (k \mp k_F)$ where
$k_F$ is the Fermi wavevector and $v_F$ is the Fermi velocity. In
order to include interactions in 1D it is often convenient to start
from a Luttinger liquid model\cite{fdmh:jpc:81} which is readily
bosonized; this approach has proved especially useful for QH edge
states. We restrict our attention here to edges of the
incompressible Hall states which occur at filling factors $\nu =1/m$
where $m$ is an odd integer. In this case it follows from
microscopic theory\cite{ahm:braz:96} that, at least for the case of
abrupt edges, edge states are described at low energies and long
wavelengths by chiral Luttinger liquid (CLL)
models\cite{wen:prb:90,wen:prb:91a,wen:int:92,mpaf:prb:92,
naga:prb:93,moon:prl:93,ludw:prl:95} with a single branch of
unidirectional boson modes. For short-range interactions between the
Fermions, CLL theory predicts non-Fermi-liquid effects for $m \ne 1$
which have been confirmed by recent experimental 
studies\cite{chang,webb:ssc:96}. The non-Fermi-liquid effects result
from vanishing weight in the quasiparticle peak in the spectral
function and {\it not} from broadening of this peak. However, to
describe 1D electron systems realistically it is often necessary to
account for the long range of the Coulomb interaction. The
importance of long range interactions for various physical
properties has been emphasized in a series of recent papers
addressing both conventional 1D electron 
systems\cite{hjs:prl:93,fab:prl:94,gia:prb:95} and QH edge state
systems\cite{oreg:prl:95,brey:95,newmoon}. In this paper we examine
the influence of the long range of the Coulomb interaction on the
spectral functions for QH edge states. 

The CLL theory long-wavelength, low-energy Hamiltonian for a $\nu =
1/m$ QH edge is\cite{wen:prb:90,wen:prb:91a,wen:int:92}:
\begin{equation}
H = \sum_{q > 0} E(q) a^{\dagger}_q a_q.
\label{eq:ham}
\end{equation} 
where $a^{\dagger}_q$ and $a_q$ are creation and annihilation
operators for the edge bosons and $E(q)$ is the mode energy. We will
ignore inter-edge interactions and for definiteness assume a disk
geometry so that $q=Q/R$ where $R$ is the radius of the disk and $Q$
is a positive integer. For interacting electron systems, these
bosons are known as edge magnetoplasmons and have the following
dispersion relation: 
\begin{equation}\label{dispersion}
E(q) = - \frac{e^2}{\epsilon \pi}\, \nu \, q \ln{(\alpha \, q \ell)}
\end{equation}
where $\alpha$ is a constant which depends on details of the
distribution of neutralizing charges which stabilize the 2DES.
This equation\cite{uz-ahm} has a broader range of validity than the
CLL model of QH edges, can be derived
classically\cite{vav-sam:jetp:88}, and has been used successfully to
interpret
experiments\cite{hls:prb:83,heit:prl:90,wass:prb:90,heit:prl:91,
ray:prb:92,hel1:85,hel2:85,hel:91} both in both QH and weaker field
regimes. For the case of coplanar neutralizing charges, $\alpha =
1/3$ for the disk\cite{pit:prl:94} geometry. The symbol $\ell:=\sqrt{
\hbar c / |e B|}$ denotes the magnetic length.

The CLL theory of QH edges is built on a powerful {\it ansatz} which
expresses the low-energy projection of the electron field operator
in terms of boson field 
operators\cite{wen:prb:90,wen:prb:91a,wen:int:92}:
\begin{mathletters}
\begin{eqnarray}\label{create}
\psi^+ (x, t) &=& \left( \frac{z}{2 \pi R} \right)^{\frac{1}{2}}
\exp{(-i [ k_F x - \mu t / \hbar ] ) } \nonumber \\ && \qquad
\qquad \times \, \exp{ ( \phi_+ (x, t) - \phi_- (x, t) \, )}  \\
\phi_+ (x, t) &=& \frac{1}{(\nu R)^{1/2}} \sum_{q>0}\frac{\exp{ 
(-i[q x - E(q) t / \hbar]) } }{q^{1/2}} \, a_{q}^{\dagger}
\end{eqnarray}
\end{mathletters}
where $\mu$ is the electron chemical potential, $z$ is a constant,
and $\phi_- (x, t) = [\phi_+ (x, t)]^+$. Eq.~(\ref{create}) is
strongly supported by recent numerical studies\cite{jujo:prl:96}.

In this work, we use expression Eq.~(\ref{create}) to evaluate the
real-time Green's function defined by  
\begin{equation}
G^>(x, t) = - i \, \left< \psi(x, t) \, \psi^+ (0, 0) \right>
\end{equation}
at zero temperature. After a standard elementary
calculation\cite{mahan} we find that 
\begin{eqnarray}\label{gpart}
&& G^>(x, t) = -i \frac{z}{2 \pi R} \nonumber \\ && \times
\exp{ \left[ \frac{1}{\nu R} \sum_{q>0} \frac{\exp{i[ (q + k_F) x -
(E(q)+\mu) t / \hbar]}}{q} \right] }  \mbox{ .}
\end{eqnarray}
The spectral function $A(k, \epsilon) \equiv A(k_F + q, \mu + \xi)$
can be written as
\begin{eqnarray}
A(k_F + q, &\mu& + \xi) = \nonumber \\ {\cal A}(Q, &\xi&) \,\,
\Theta(q) \,\Theta(\xi) + {\cal A}(-Q, -\xi) \,\, \Theta(- q) \,
\Theta(- \xi)
\end{eqnarray}
with the function ${\cal A}(Q, \xi) = - \frac{1}{2\pi} \Im m
\{ G^{>}(k_F + q, \mu + \xi)\}$. We evaluate ${\cal A}(Q, \xi)$ by
taking Fourier transforms of (\ref{gpart}) with respect to $x$ and
$t$. Here the integer $Q \equiv q R > 0$ is the wavevector measured
from the Fermi wavevector in units of $R^{-1}$.

In the short-range interaction case, $E(q) = \hbar c q$ so that edge
phonons propagate without dispersion at velocity $c$ and the
spectral function has a $\delta$-function peak at $\xi=\xi_k \equiv
\hbar c (k-k_F)$.
\begin{figure}[b]
\epsfxsize=3.8in
\centerline{\epsffile{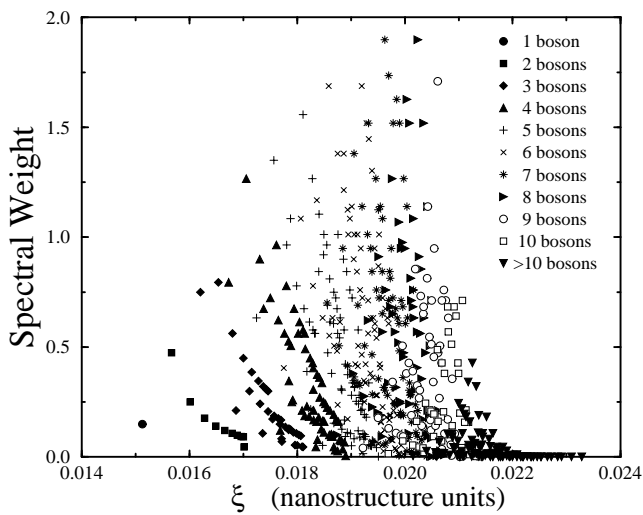}}
\caption{Structure in the spectral function (\protect\ref{spectral})
for fixed $Q=20$. Shown are the weights of the peaks in ${\cal A}(Q,
\xi)$ as a function of the energy $\xi$ (unit: $\frac{e^2}{\epsilon
\ell\pi}\sim 4.4\,$meV at 10 T) at which they are located. Data points
denoted by the same symbol represent states with the same total number
of bosons $\aleph$ in the system. Indicated are peaks for $\aleph = 1,
2, \dots, 10$. These results are for $\nu = 1/3$ and $R=2500 \ell$.
At typical fields this corresponds to $R \sim 20\mu$m.}
\label{peakbunch}
\end{figure}
For the realistic long-range interaction case,
however, it is necessary to expand the exponential in
Eq.~(\ref{gpart}) before Fourier transforming.
For a given value of
$Q$ we obtain a separate delta function in energy for each way in
which the integer $Q$ can be expressed as a sum of integers, thus
relating the spectral function calculation to the mathematical
theory of partitions\cite{math2:76}. A partition of an integer $Q$
is a decomposition of $Q$ into the sum of integers $n\le Q$: $Q =
\sum_{n} n \, l_n$ where each integer $n$ occurs $l_n$ times.
Partitions are usually specified by the symbol $(l) \equiv ( 1^{l_1}
2^{l_2} \dots Q^{l_Q} )$. Using this notation, we can write
\begin{mathletters}
\begin{equation}\label{spectral}
{\cal A}(Q, \xi) = z \sum_{(l) \mbox{\tiny  ~of } Q} w^{(l)}_{\nu}
\,\, \delta \big(\xi - E^{(l)} \big)
\end{equation}
where the sum is taken over all partitions $(l)$ of $Q$, and the
expressions for the weights $w^{(l)}_{\nu}$ and energies $E^{(l)}$
are
\begin{equation}\label{weight}
w^{(l)}_{\nu} = \left[ \prod_{n\in (l)} (\nu n)^{l_n} l_n !
\right]^{-1}
\end{equation}
and
\begin{equation}\label{energies}
E^{(l)} = \sum_{n\in (l)} l_n \, E(n/R)
\end{equation}
respectively.
\end{mathletters}
For small $Q$ all partitions are readily enumerated and expressions 
(\ref{weight}) and (\ref{energies}) evaluated. In Fig.\
\ref{peakbunch} we show the 627 peaks which occur in the spectral
function for $Q=20$.

The formal structure of expression (\ref{spectral}) can be
understood by noting that, for fixed $Q$, each partition $(l)$ is
uniquely related to a microscopic many-particle state. The strength
$w^{(l)}_{\nu}$ of the contribution of a particular state
represented by the partition $(l)$ to the spectral function is
proportional to the probability that the state created by adding an
electron to the edge of a QH system will have total excess momentum
$Q/R$ and $l_n$ bosons with momentum $n/R$. The excitation with the
smallest energy possible at wavevector $q=Q/R$ gives rise to a
$\delta$-function peak at $\xi = \xi^{\mbox{\tiny min}}_q = E(q)$
and corresponds to the state with exactly one boson with momentum
$q$ in the system (cf.\ left-most symbol [filled circle] in
Fig.~\ref{peakbunch}). The largest excitation energy
$\xi^{\mbox{\tiny max}}_q$ occurs for the state with $Q$ bosons of
momentum $1/R$, represented by the right-most symbol (triangle down)
in Fig.~\ref{peakbunch}:
\begin{equation}
\xi^{\mbox{\tiny max}}_{q} = E(q) \left[ 1 + \frac{\ln(q R)}{
\ln(1/\alpha \, q \ell)} \right] \mbox{ .}
\end{equation}
The total number of bosons for a state represented by partition
$(l)$ is $\aleph = \sum_{n\in (l)} l_n$. In Fig.\ \ref{peakbunch},
data points with the same symbol correspond to states with the same
total boson number $\aleph$. The sum of all their weights is known
from number theory\cite{math2:76}; it can be expressed in terms of the
so-called {\it Stirling numbers of the first kind}\cite{math2:76}
which are usually denoted by $S_{Q}^{(\aleph)}$. We observe that the
peaks with large weight for states with fixed $\aleph$ cluster near
the same energy $\xi_\aleph$. Furthermore, it is apparent from the
figure (and easily understood formally) that the spectral weight of
states with large $\aleph$ is small. However, the number of these
states grows rapidly with $Q$. In order to estimate the line shape
of the spectral function, we have to consider both the magnitude of
the spectral weights for different states as well as the number of
states available within a particular energy interval. We used the
Stirling formula to approximately determine the partition
$(l)_{\aleph}$ which maximizes the weight (Eq.(\ref{weight})) for
fixed $\aleph$. Since most of the peaks for fixed $\aleph$ cluster
near $E^{(l)_{\aleph}}$, we propose the following approximate
formula for the spectral function: 
\begin{equation}\label{approx}
\tilde{\cal A}(Q, \xi) = z \sum_{\aleph=1}^{Q} \nu^{-\aleph} 
\frac{|S_{Q}^{(\aleph)}|}{Q!} \, \, \delta \big(\xi -
E^{(l)_{\aleph}} \big) \mbox{ .}
\end{equation}
Eq.~(\ref{approx}) can be used to determine the energy at which the
spectral function is peaked (for $q R > \nu^{-2}$):
\begin{equation}\label{peakeng}
\xi^{\mbox{\tiny peak}}_q \sim E(q) \left[ 1 + \frac{\nu^{-2}}{q R}
\, \frac{\ln(q R)}{\ln(1/\alpha \, q \ell)} \right] \mbox{ .}
\end{equation}
In the case of short-range interaction, the effect of a fractional
filling factor $\nu$ is to shift spectral weight towards higher
energies. If the interaction is long-ranged, this effect is
enhanced, and the suppression of spectral weight at low energies is
effectively stronger. As the comparison of Eq.~(\ref{peakeng}) and
$\xi^{\mbox{\tiny min}},\xi^{\mbox{\tiny max}}$ shows, this
enhancement is weakened at large $Q=qR$.

Recently, Luttinger liquid behavior of QH edges was observed in
measurements of the IV-characteristics for tunneling from a
reservoir into QH edges\cite{chang} and for tunneling between two
co-planar edge channels\cite{webb:ssc:96}.
\begin{figure}[b]
\epsfxsize3.4in
\centerline{\epsffile{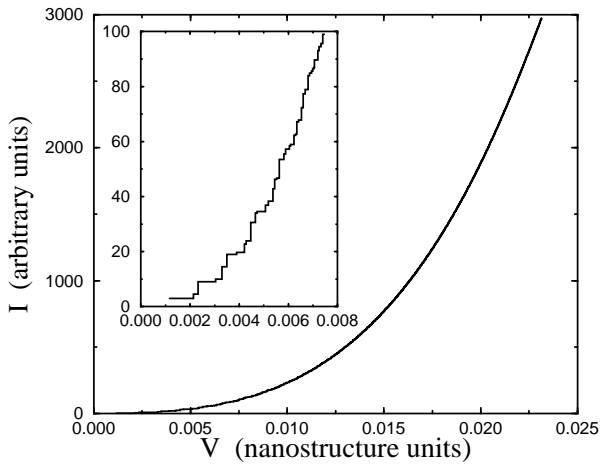}}
\caption{IV-curve for tunneling from a particle reservoir into a QH
edge ($\nu = 1/3$, $R=2500\ell \sim 20\mu$m typically, corresponding
to $10^6$ electrons). Inset shows step-structure resulting from
finite size of the system. Step width in the voltage is non-uniform,
which is a signature of Coulomb interaction. Nanostructure units:
$[V] = \frac{e}{\epsilon \ell \pi} \sim 4.4$mV here.}
\label{ivfig}
\end{figure}
In both experiments,
translational symmetry was broken, and the tunneling process was not
restricted by momentum conservation. We will show below that in
experiments of this kind, the broadening of the spectral function
due to the long range of Coulomb interaction (as discussed above)
does not lead to easily observable effects. Instead, we propose an
experiment which measures tunneling currents between two-dimensional
electron systems\cite{weim:apl:88,eis:prb:91} displaced along their
normals. This so-called {\it vertical tunneling} experiment is
directly sensitive to the line shape of the spectral function.

According to standard theory\cite{mahan} based on the tunneling
Hamiltonian formalism, the expression for the IV-characteristics for
tunneling between two systems (labelled by indices $U,L$,
respectively) which are separated by a barrier is
\begin{eqnarray}\label{tunnham}
&& I(V) = \nonumber \\ && 2\pi e \sum_{k, p} \big| t_{k p} \big|^2 
\int_{\mu^{(U)} - e V}^{\mu^{(L)}} d\epsilon \,\, A^{(L)}(k,
\epsilon) A^{(U)}(p, \epsilon + e V)
\end{eqnarray}
at zero temperature. If we consider tunneling from a reservoir into
a QH edge, the tunneling amplitude will not importantly depend on
momentum: $\big| t_{k p} \big|^2 \sim t^2$. Eq.~(\ref{tunnham}) then
specializes to
\begin{equation}
I(V) = 2 \pi e t^2 \, {\cal N} \, \sum_{Q} \int_{0}^{e V}
d\xi \,\,\, {\cal A}(Q, \xi)
\end{equation}
where we assumed the reservoir's DOS to be essentially constant
($\equiv {\cal N}$) and the chemical potential in both systems to be
equal. As ${\cal A}(Q, \xi)$ is a sum of $\delta$-functions in the
variable $\xi$, the integration is easily performed. The
numerically-determined IV-curve is shown in Fig.\ \ref{ivfig}. It
turns out that, for experimentally realistic sample sizes, the
IV-curve shows power-law behavior $I \sim V^{1/\nu}$ as 
predicted\cite{wen:int:92} for the short-range interaction case. The
long range of Coulomb interaction manifests itself only in small
corrections (logarithmic in the voltage) to the power-law. The
reason is that a tunneling probe {\it within} the plane of the 2DES
is sensitive only to the total spectral weight for energies below
$eV$, and the width of the spectral functions becomes unimportant.
Note that the smallest possible excitation energy ({\it i.e.}
$E(1/R)$ ) is quite big even for large samples; typical numbers are
$4.6\, \mu$eV ($0.14\, \mu$eV) for $R = 20\, \mu$m (1mm) at 10T and
$\nu = 1/3$. This leads to the appreciable finite-size effects seen
in the inset of Fig.~\ref{ivfig} at low voltages. For a short-range
interaction, the step-size in voltage would be uniform. It is a
signature of the Coulomb interaction that the voltage step-size is
irregular.

The situation is different if we consider tunneling between two QH
edges which are separated vertically, {\it i.e.} belong to two
different 2DES which are parallel to each other. As tunneling occurs
now from/into the whole (1D) edges, momentum conservation severely
restricts the tunneling process, and we have $\big| t_{k p} \big|^2
\longrightarrow t^2 \, \delta_{kp}$. For tunneling to occur, we have
to create an excitation with momentum below the Fermi point in one
system and an excitation above the Fermi point in the other at the
same time. Therefore, in a {\it vertical tunneling} experiment, a
nonzero current will occur at finite voltages only if the Fermi
wavevectors $k_F^{(U)}, k_F^{(L)}$ in the two systems are different:
$(k_F^{(L)} - k_F^{(U)}) R \equiv \tilde Q > 0$. In other words, if
the Fermi wavevectors in the systems $U,L$ are different,
$k$-conservation does not mean $q$-conservation (remember: $k\equiv
k_F + q$). Expression (\ref{tunnham}) can be evaluated analytically
for {\it vertical tunneling} and yields
\begin{equation}\label{vertun}
I(V) = 2\pi e t^2 z^2 \sum_{(l) \mbox{\tiny ~of } \tilde Q}
w^{(l)}_{\nu/2} \,\, \delta \big(eV - E^{(l)}\big) \mbox{ .}
\end{equation}
This result means that, by tuning\cite{comm1} the off-set in the
Fermi wavevectors for the two vertically-separated edges, it is
possible to explore the structure in the spectral function
{\it directly} by measuring the tunneling IV-curve. The tunneling
current (\ref{vertun}) has peaks at voltages corresponding to the
peaks in ${\cal A}(\tilde Q, \xi)$. The weight of the peaks in the
the IV-curve is simply related to the spectral weight of the 
corresponding peak in ${\cal A}(\tilde Q, \xi)$. Hence the IV-curve
for fixed $\tilde Q$ looks similar to Fig.~\ref{peakbunch}. However,
peaks at higher voltages will be enhanced even more strongly. In
experiment, the broadening of each peak (due to electron-phonon
coupling and disorder effects) has to be taken into account.

In summary, we have calculated the spectral function for electrons
at quantum Hall edges with Coulomb interaction present. We quantify
its finite width, and give expressions for tunneling IV-curves.
{\it Vertical tunneling} provides a direct probe of the line shape.

The authors would like to thank S.M.\ Girvin, J.J.\ Palacios, and
R.\ Haussmann for numerous useful discussions. This work was funded
in part by the National Science Foundation under Grant No.\
DMR-9416906. U.Z.\ gratefully acknowledges financial support from
Studien\-stiftung des deutschen Volkes (Bonn, Germany).


\begin{thebibliography}{10}

\bibitem{ahmintro}
A.~H. MacDonald,  in {\em Proceedings of the 1994 Les Houches Summer
School on Mesoscopic Physics}, edited by E. Akkermans {\it et~al.}
(Elsevier Science, Amsterdam, 1995), pp.\ 659--720.

\bibitem{rbl:prb:81}
R.~B. Laughlin, Phys. Rev. B {\bf 23},  5632  (1981).

\bibitem{bih:prb:82}
B.~I. Halperin, Phys. Rev. B {\bf 25},  2185  (1982).

\bibitem{pav:prb:84}
A.~H. MacDonald and P. St{\v r}eda, Phys. Rev. B {\bf 29},  1616
(1984).

\bibitem{butt:prb:88}
M. B\"uttiker, Phys. Rev. B {\bf 38},  9375  (1988).

\bibitem{emery}
V.~J. Emery,  in {\em Highly Conducting One-Dimensional Solids},
edited by J.~T. Devreese {\it et~al.} (Plenum Press, New York,
1979), pp.\ 247--303.

\bibitem{sol:adv:79}
J. S\a'olyom, Adv. Phys. {\bf 28},  201  (1979).

\bibitem{fdmh:jpc:81}
F.~D.~M. Haldane, J. Phys. C {\bf 14},  2585  (1981).

\bibitem{apel:prb:82}
W. Apel and T.~M. Rice, Phys. Rev. B {\bf 26},  7063  (1982).

\bibitem{ahm:braz:96}
A.~H. MacDonald, Brazilian J. Phys. {\bf 26},  43  (1996).

\bibitem{wen:prb:90}
X.~G. Wen, Phys. Rev. B {\bf 41},  12838  (1990).

\bibitem{wen:prb:91a}
X.~G. Wen, Phys. Rev. B {\bf 44},  5708  (1991).

\bibitem{wen:int:92}
X.~G. Wen, Int. Journ. Mod. Phys. B {\bf 6},  1711  (1992).

\bibitem{mpaf:prb:92}
C.~L. Kane and M.~P.~A. Fisher, Phys. Rev. B {\bf 46},  15233
(1992).

\bibitem{naga:prb:93}
A. Furusaki and N. Nagaosa, Phys. Rev. B {\bf 47},  3827  (1993).

\bibitem{moon:prl:93}
K. Moon {\it et~al.}, Phys. Rev. Lett. {\bf 71},  4381  (1993).

\bibitem{ludw:prl:95}
P. Fendley, A.~W.~W. Ludwig, and H. Saleur, Phys. Rev. Lett.
{\bf 74},  3005 (1995).

\bibitem{chang}
A.~M. Chang, L.~N. Pfeiffer, and K.~W. West, Phys. Rev. Lett. {\bf 77}
(to appear).

\bibitem{webb:ssc:96}
F.~P. Milliken, C.~P. Umbach, and R.~A. Webb, Solid State Comm.
{\bf 97},  309 (1996).

\bibitem{hjs:prl:93}
H.~J. Schulz, Phys. Rev. Lett. {\bf 71},  1864  (1993).

\bibitem{fab:prl:94}
M. Fabrizio, A.~O. Gogolin, and S. Scheidl, Phys. Rev. Lett.
{\bf 72},  2235 (1994).

\bibitem{gia:prb:95}
H. Maurey and T. Giamarchi, Phys. Rev. B {\bf 51},  10833  (1995).

\bibitem{oreg:prl:95}
Y. Oreg and A.~M. Finkel'stein, Phys. Rev. Lett. {\bf 74},  3668
(1995).

\bibitem{brey:95}
M. Franco and L. Brey, Phys. Rev. Lett. {\bf 77}, 1358 (1996).

\bibitem{newmoon}
K. Moon and S.~M. Girvin, Phys. Rev. B {\bf 54}, 4448 (1996).

\bibitem{uz-ahm}
U. Z\"ulicke and A.~H. MacDonald (unpublished).

\bibitem{vav-sam:jetp:88}
V.~A. Volkov and S.~A. Mikhailov, Zh. Eksp. Teor. Fiz. {\bf 94},
217 (1988) [Sov. Phys. JETP {\bf 67},  1639 (1988)].

\bibitem{hls:prb:83}
S.~J. Allen, H.~L. St\"ormer, and J.~C.~M. Hwang, Phys. Rev. B
{\bf 28},  4875 (1983).

\bibitem{heit:prl:90}
T. Demel, D. Heitmann, P. Grambow, and K. Ploog, Phys. Rev. Lett.
{\bf 64}, 788  (1990).

\bibitem{wass:prb:90}
M. Wassermeier {\it et~al.}, Phys. Rev. B {\bf 41},  10287  (1990).

\bibitem{heit:prl:91}
I. Grodnensky, D. Heitmann, and K. von Klitzing, Phys. Rev. Lett.
{\bf 67}, 1091  (1991).

\bibitem{ray:prb:92}
R.~C. Ashoori {\it et~al.}, Phys. Rev. B {\bf 45},  3894  (1992).

\bibitem{hel1:85}
D.~B. Mast, A.~J. Dahm, and A.~L. Fetter, Phys. Rev. Lett {\bf 54},
1706 (1985).

\bibitem{hel2:85}
D.~C. Glattli {\it et~al.}, Phys. Rev. Lett {\bf 54},  1710  (1985).

\bibitem{hel:91}
P.~J.~M. Peters {\it et~al.}, Phys. Rev. Lett. {\bf 67},  2199
(1991).

\bibitem{pit:prl:94}
S. Giovanazzi, L. Pitaevskii, and S. Stringari, Phys. Rev Lett
{\bf 72},  3230 (1994).

\bibitem{jujo:prl:96}
J.~J. Palacios and A.~H. MacDonald, Phys. Rev. Lett. {\bf 76},  118
(1996).

\bibitem{mahan}
G.~D. Mahan, {\em Many-Particle Physics} (Plenum, New York, 1990).

\bibitem{math2:76}
G.~E. Andrews, {\em The Theory of Partitions} (Addison-Wesley,
Reading, 1976); see also K. Sch\"onhammer and V. Meden, Phys. Rev.
B {\bf 47}, 16205 (1993).

\bibitem{weim:apl:88}
J. Smoliner, E. Gornik, and G. Weimann, Appl. Phys. Lett. {\bf 52},
2136 (1988).

\bibitem{eis:prb:91}
J.~P. Eisenstein, T.~J. Gramila, L.~N. Pfeiffer, and K.~W. West,
Phys. Rev. B {\bf 44},  6511  (1991).

\bibitem{comm1}
A possible way to tune $\tilde Q$ is to apply an in-plane magnetic
field. By measuring the tunneling current, the experiment can be
calibrated. At $\tilde Q = 0$, the only peak in the IV-curve will
occur at zero voltage. Having achieved this calibration, it is
possible to set the in-plane magnetic field for accurately
determining $\tilde Q$.

\end{thebibliography}
\end{document}